\documentclass[twocolumn,showpacs,preprintnumbers,amsmath,amssymb,revsymb,prb, superscriptaddress]{revtex4}
\usepackage{graphicx}
\usepackage{mathptmx}
\usepackage{color}

\newcommand{\D}[2]{\frac{d #1}{d #2}}

\newcommand{\AF}{ }

\begin{document}
\title{Tunneling through Al/AlOx/Al junction: analytical models and first principles simulations}
\author{M. Zemanov\'{a} Die\v{s}kov\'{a}}
\affiliation{Department of Physics, Faculty of Electrical Engineering and Information Technology, 
	Slovak University of Technology, Ilkovi\v{c}ova 3, 812 19 Bratislava, Slovak Republic}
\author{A. Ferretti}
\affiliation{\AF Centro S3, CNR--Istituto Nanoscienze, I-41125 Modena, Italy.}
\affiliation{European Theoretical Spectroscopy Facility (ETSF, {\tt www.etsf.eu})}
\author{P. Bokes } \email{peter.bokes@stuba.sk}
\affiliation{Department of Physics, Faculty of Electrical Engineering and
        Information Technology, Slovak University of Technology,
    Ilkovi\v{c}ova 3, 812 19 Bratislava, Slovak Republic}
\affiliation{European Theoretical Spectroscopy Facility (ETSF, {\tt www.etsf.eu})}

\date{\today{}}

\begin{abstract}
{\AF
We study from first principles the transport properties of Al/AlO$_x$/Al tunnel junctions.
On this basis, we analyze the reliability of two analytical models for the conductance, namely the trapezoid 
potential barrier model and a tight-binding model.
Our findings show that ($i$) the interface width used in the models is determined by the 
electronic density profile, and it is shorter than the width one expects from the atomic arrangements;
($ii$) the effective mass}, found to be about 
on third of the free electron mass, can be determined from the oxide band-structure calculations, and 
($iii$) the barrier height is given by one fourth of the bandgap in the oxide, 
{\AF which explains the apparently small values found for } these junctions experimentally.
\end{abstract}


\maketitle

\section{Introduction}
\label{sec-1}
Tunneling of electrons through aluminum-aluminum oxide (Al/AlO$_x$/Al) 
junctions is one of the prototypical examples of quantum-mechanical tunneling in solid state physics. 
Fisher and Giaever in their pioneering work~\cite{Fisher1961} demonstrated the tunneling character 
of the transport of electrons through this interface and by comparing their results with 
the predictions of Holm~\cite{Holm1951} for tunneling through vacuum gap, they
initiated the interpretation of tunneling measurements through thin metal-insulator-metal 
junctions using potential barrier model. The minimal form of this model contains two parameters - 
the barrier width $d$, which indicates the physical width of the oxide, and its height $W$, {\AF given by the}
energy difference between the Fermi energy and the bottom of the conduction band in the oxide.
In practice, several other parameters enter the model~\cite{Stratton62,Simmons1963,Brinkman1970}: 
the electron's effective mass in the oxide or the dielectric constant of the oxide used within 
{\AF an} additional image-charge potential. Further parameters are used for fine-tuning the shape of 
the barrier, e.g.  it's asymmetry~\cite{Brinkman1970}. Clearly, having a large set of parameters, 
it is no surprise that the simple barrier model can be fitted to the experimental current-voltage 
characteristics well~\cite{Groner2002,Buchanan2002,Gloos03,Schaefer2011}, but at the same time, 
it rises questions about the relevance of the model {\AF itself.}~\cite{Miller2007,Lacquaniti2012}
For example, the inclusion of the image potential can have a significant effect on the effective 
barrier width, but its presence depends on the time scales of the tunneling electrons and 
the interface plasmons in the metal~\cite{Jonson80,Bokes11}.

On the other hand, much more detailed and parameter-free models of the interface can be constructed 
using first principles calculations~\cite{Feibelman07,Jung09,Fadlallah09,Stoeffler02,
Belashchenko05,Zhuravlev11} even though the size of the modeled interfaces is somewhat restricted 
due to {\AF the} numerical cost of these calculations. {\AF Nevertheless}, 
in many experiments~\cite{Rippard02,Gloos03,Tan05,Koppinen2007,Jung09,Schaefer2011,Lacquaniti2012} 
the studied interfaces 
have widths within the reach of {\it ab initio} simulations so {\AF that 
the accuracy} of the potential barrier model to the interpretation {\AF of tunneling data} can be tested. 
Specifically, Jung {\it et al}~\cite{Jung09} presented such a study comparing the character 
of the equilibrium projected density of states of the Al/AlO$_x$/Al interface obtained by 
a first principles simulation with the potential barrier model. {\AF They} found that 
the parameters of the potential barrier model fitted to the experimental data are in qualitative agreement 
with the parameters of the first principles {\AF calculations}. The potential barrier model included the image potential 
and hence also the dielectric constant which effectively narrowed and lowered the potential barrier.

In {\AF this} work we test the performance of the potential barrier model by comparing {\AF the predicted}
conductance to {\it ab initio} {\AF calculations}~\cite{Smogunov2004, Ferretti05, Ferretti2007}.
{\AF We test} this for Al/AlO$_x$/Al junctions of four different widths $d$ and show that  
it is essential to use {\AF an} effective mass in the oxide and {\AF an} effectively shorter width of the 
{\AF tunneling region} within the potential barrier model. We also present an analytical tight-binding model 
for the conductance that {\AF describes} the {\it ab initio} {\AF results more accurately} than 
the potential barrier {\AF model. The parameters of the latter are extracted from} the ground state 
{\it ab initio} calculations of the junction. 
{\AF In Sec.~{\ref{sec-2}} and {\ref{sec-3}} we introduce the analytical details of the models.
The {\it ab initio} results for ground state properties of the studied junctions are presented 
in Sec.~{\ref{sec-4}} and {\ref{sec-4b}}, together with the computational parameters used in the calculations. 
Finally in Sec.~{\ref{sec-5}} we compare the conductances obtained using the {\it ab initio} calculations
and the conductances obained from the analyical models.
}

\section{Potential barrier models of the interface}
\label{sec-2} 

The starting assumption of the potential barrier model is that inside the metallic electrodes,
on the left and right of the insulator, the electrons behave like free quasi-particles 
with their energy being in a separable form~\footnote{{\AF Throughout the paper we use Hartree atomic units:
Energies are in Hartree ($Ha=27.2$eV) and lengths in units of the Bohr radius $a_B=0.529$\AA.}}
\begin{equation}
	E = E_z + E_\parallel = k_z^2/2 + k_\parallel^2/2
\end{equation}
where $k_z$ is the component of electron's momentum perpendicular to the interface and $k_\parallel$ {\AF the 
component of momentum} parallel to the interface.
The current density, induced by an infinitesimal bias voltage, consists of a sum of contributions 
from the electrons occupying states in the energy window around 
the Fermi energy $E_F$, with their momentum opposite to the drop of the bias voltage ($k_z>0$). Hence,
the conductance per area is given by 
the {\AF expression}~\footnote{In our atomic units, the conductance per unit area
is given in {\AF units} of $e^2 a_B^{-2}/\hbar = 8.692 10^{10} \mu$S$\mu$m$^{-2}$.}
\begin{eqnarray}
	g &=& 2 \int \frac{d^2k_{\parallel}}{(2\pi)^2} 
		\int_{-\infty}^{\infty} \frac{dE_z}{2\pi}
                \delta(E_F - E_z - E_\parallel) T(E_z), \label{eq-2-1} \\
	&=& \int_{0}^{\infty} \frac{dE_\parallel}{2\pi^2} T(E_F-E_\parallel), \label{eq-2-2}
\end{eqnarray}
where $T(E_z)$ is the transmission function, the probability for an electron to pass through the junction, 
and $E_F$ is the Fermi energy. 

\begin{figure}[!t]
\begin{center}
       \includegraphics[width=7cm]{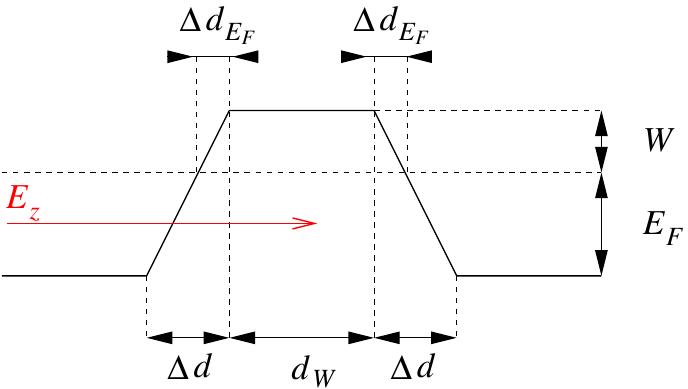}
\caption{(Color online) The trapezoid potential barrier used to model ultra-thin AlO$_X$ interface.
	} \label{fig-2-1}
\end{center}
\end{figure}

The simplest expression for the transmission $T(E_z)$ is based on a metal-vacuum-metal 
interface~\cite{Stratton62, Brinkman1970, Simmons1963}, where {\AF the barrier height $W$ is given by 
the potential energy in the vacuum with respect to the Fermi energy of the metal.}

In Sec.~\ref{sec-5} we will demonstrate that there are two essential features of the potential 
barrier models that need to be taken into account for the description 
of ultra-thin interfaces: (1) the barrier needs to have transition regions between the metal 
and the insulator of width $\Delta d$, where the potential energy changes continuously, 
(2) the effective mass of the electrons in the insulator needs to be accounted for. 
These two requirements can be fulfilled by using a specific {\AF shape of the potential barrier. 
In this work we use a trapezoid potential barrier (TB) as defined in Fig.~\ref{fig-2-1}.}

On the other hand, in Sec.~\ref{sec-5} we will also demonstrate that {\AF
approximate expressions for the transmission as well as for the energy 
integration in Eq.~(\ref{eq-2-2}) are sufficient for an accurate evaluation of the model conductance}. 
For the trapezoid potential barrier{\AF,} the WKB approximation for the transmission gives 
\begin{eqnarray} 
	T(E - E_\parallel) &=& \exp\left\{-F(E - E_\parallel)\right\}, \label{eq-2-3} \\
	F(E - E_\parallel) &=& 2 \int_{-d(E_\parallel)}^{d_(E_\parallel)} 
	\sqrt{2 m_{\AF \text{eff}} [W(z)+E_\parallel]} dz, \label{eq-2-4}
\end{eqnarray}
where $m_{\AF\text{eff}}$ is the effective mass of the electrons in the insulator, $-d(E_\parallel)$ 
and $d(E_\parallel)$  
give the region where $\AF[W(z)+E_\parallel]\geq 0$, and $W(z)$ is the trapezoid potential profile.
Accounting only for the largest contribution from the states close to the Fermi energy in the integral
in {\AF Eq.~(\ref{eq-2-2})}, $E_\parallel \sim E_F$, we obtain {\AF the following simple anaytical expression:}
\begin{equation}
	g \approx - \frac{e^{-F(E_F)}}{2\pi^2 F'(E_F)}, \label{eq-2-6}
\end{equation}
where 
\begin{eqnarray}
	F(E_F) &=& 2 \sqrt{2m_{\AF\text{eff}}W} \left( d_W + \frac{2}{3} \Delta d_{E_F} \right),  \label{eq-2-7} \\
	F'(E_F) &=& - \frac{2}{\sqrt{2 m_{\AF\text{eff}} W}} \left( d_W + 2 \Delta d_{E_F} \right).
\end{eqnarray}
We will refer {\AF to Eq.~(\ref{eq-2-6})} as the TB$^{A,m_{\AF\text{eff}}}$ model ($A$ {\AF stands} 
for ``analytical'' as compared 
to the numerically calculated transmission for the trapezoid potential barrier - TB$^{N}$). 
{\AF We note that the introducion of two transition regions of width $\Delta d$
adds to the exponent of the transmission amplitude only a small fraction of $\Delta d$, namely $(2/3)\Delta d_{E_F}$.}
This results {\AF in a} substantial increase of the conductance which {\AF is needed for the agreement 
of the TB model and {\it ab initio} results (see Sec.~\ref{sec-5}).}

\section{Atomic $sp$ model of the insulator}
\label{sec-3} 

It is typically assumed that the barrier height in the potential barrier model corresponds to 
energy distance between the Fermi energy and the {\AF closest among the} valence or conduction bands
of the insulator, or even to its whole bandgap. However, fits of the potential 
barrier model to experimental data often lead to unphysically small values if one {\AF follows} this interpretation. 
Various arguments like interface 
roughness~\cite{Miller2007} or image potential~\cite{Simmons1963} have been suggested to correct 
for this {\AF underestimation}, but perhaps the most important one -- the principal difference in the energetic spectrum 
of the real insulator and the vacuum gap {\AF --} received less attention~\cite{Stratton62,Gundlach1973}.

To account for a more realistic electronic structure of the insulator 
we consider a minimal tight-binding model of a $sp$-like insulator with rock-salt crystal structure. 
For our purposes, the cation with $s$-like orbital plays the role of {\AF aluminum} and the anion 
with $p$-like orbital the oxygen {\AF atom}. While this is different from the true structure 
{\AF of alumina}, this model works surprisingly well even for the disordered aluminum oxide found 
in our interfaces, as will be shown in Sec.~\ref{sec-4}.

{\AF The $sp$ model has four parameters: the onsite} atomic energies of the cation ($\epsilon_s$) 
and anion ($\epsilon_p$), the hopping matrix element between the two atoms ($t$), and the length 
of the edge of the conventional unit cell (cube) $a$. A standard calculation leads 
to valence (v) and conduction (c) {\AF band energies}
\begin{equation}
        \AF
	E_{\textrm{c}/\textrm{v}}(\mathbf{k}) = E_F^{\infty} \pm \frac{E_g}{2}\,
		\sqrt{ 
                1 + \frac{8 m_{\text{eff}}^{-1}}{E_g a^2} 
		\sum_{i=1}^{3}\sin^2(k_i a/2) \,
                }
        \label{eq-3-1}
\end{equation}
where $m_{\AF\text{eff}}=E_g/(2 t^2 a^2)$ is the effective mass of the {\AF electrons close 
to the conduction band minimum, equal in magnitude that 
of the valence band maximum}. The two bands are separated by the bandgap 
	$\AF E_g = \epsilon_p - \epsilon_s$,
and the energy in the middle of the gap is
\begin{equation}
    \AF
	E^{\infty}_F = \frac{\epsilon_p + \epsilon_s}{2}.
\end{equation}

In the tunneling regime, {\AF the current is carried by the electronic states 
in the bandgap,~\cite{Tomfohr2002,Prodan07,Ferretti2012} 
i.e. the evanescent Bloch states with imaginary wavenumber $k_z = i \kappa$:
\begin{equation}
	\phi_{\kappa,k_\parallel}(\mathbf{r}) \sim 
	e^{-\kappa z} e^{i \mathbf{k}_\parallel \cdot \mathbf{r}} u_{\kappa,\mathbf{k}_\parallel}(\mathbf{r}) .
\end{equation}
The WKB-like result for the transmission takes then} the form
\begin{equation}
	T^{sp}_{k_\parallel}(E) \sim |\phi_{\kappa,k_\parallel}(\AF\mathbf{d})|^2 
              \sim e^{-2 \kappa(E,k_\parallel) d}, \label{eq-3-5}
\end{equation}
where $\AF \mathbf{d}$ is a vector normal to the interface with the length given by the width of 
the interface ($\AF |\mathbf{d}| \sim d$).
$\kappa(E,k_\parallel)$ can be obtained from {\AF Eq.~(\ref{eq-3-1})} using the substitution $k_z=i\kappa$ therein.
The transmission can be then used in the calculation of the conductance in Eq.~\ref{eq-2-1}. The largest contributions 
to the conductance come only from $\kappa a/2 < 1$, $k_\parallel a/2 <1$, so that the sin$(~)$ functions 
in the dispersion can be expanded in Taylor series. Keeping only the first two terms
we find~\footnote{The resulting expression is similar to the non-parabolic model discussed 
by Stratton~\cite{Stratton62} or Gundlach~\cite{Gundlach1973} and credited to Franz~\cite{Franz56}.}
\begin{eqnarray}
	\kappa(E,k_\parallel) &=& \sqrt{ \nu(E) m_{\AF \text{eff}} E_g /2 + k_x^2 + k_y^2 }, \\
			&=& \sqrt{ 2 \AF \left[ \nu(E) m_{\AF\text{eff}} E_g /4 + E_\parallel \right] },
\end{eqnarray}
where we have introduced a multiplicative factor $\nu(E)$ accounting for the relative distance of 
the energy $E$ from the middle of the gap,
\begin{equation}
	\nu(E) = 1 - 4 \left( \frac{E-E_F^\infty}{E_g} \right)^2, \label{eq-3-7}
\end{equation}
{\AF which is close to 1 for $E \sim E_F$. We note that} by using the Taylor expansion the model
{\AF becames} independent of the size of the conventional cell $a$.
The transmission $T^{sp}_{k_\parallel}(E)$ is similar to the WKB result 
for a potential barrier {\AF[Eqs.~(\ref{eq-2-3}-\ref{eq-2-4}) for a constant barrier height $W$]}. Hence,
making the same approximations as in Sec.~\ref{sec-2} and substituting $W \rightarrow \nu(E_F) E_g/4$ 
we find an analytical {\AF expression} for the transmission through a $sp$ insulator of width $d$ precisely of the 
form {of Eq.~(\ref{eq-2-6}),} where 
\begin{eqnarray}
	F_{sp}(E_F) &=& 2 \sqrt{\nu(E_F) m_{\AF\text{eff}} E_g /2} \ d,  \label{eq-3-8} \\
	F_{sp}'(E_F) &=& - \frac{2}{\sqrt{ \nu(E_F) m_{\AF\text{eff}} E_g/2}} \ d. \label{eq-3-9}
\end{eqnarray}
This represents one of the main results of our paper: the potential barrier height $W$ is related to the 
bandgap through the relation $W = \nu(E_F) E_g/4$. Since the Fermi energy in our junctions is close to
the center of the gap (Sec.~\ref{sec-4}) {\AF where we have} $\nu(E_F) \sim 1$, 
we expect that the bandgap is about four times
larger than the barrier height obtained from the fits to the experimental data. This explains the typical 
situation in Al/AlO$_x$/Al junctions where $W$ can be as small as $2$eV or less, which is to be compared 
with the bandgap of alumina being about $7 - 9$eV. Further comparisons will be made in 
the Sec.~\ref{sec-5} where the $sp$ model is compared to the {\it ab initio} calculation 
of the conductance.

\section{First principles calculations of the Al/AlO$_x$/Al interfaces}
\label{sec-4}

The Al/AlO$_x$ thin film is well known for its difficulties to be grown in an ordered form~\cite{Mizuguchi2005,Chen08}. 
The process of oxidation consists of a quick chemisorption of oxygen on a clean Al surface which is followed by 
a complex diffusion process leading to various widths of the interface which is typically 
disordered~\cite{Kravchuk04,Nesbitt2007,Koppinen2007,Hasnaoui05}. 
The model that we consider is on the other hand relatively simple and ordered. We followed 
Jennison~\cite{Jennison1999,Jennison2000} at constructing chemisorbed layer of oxygen on an ideal Al(111) 
{\AF $\sqrt{3} \times \sqrt{3}$ surface} (three Al atoms per layer), {\AF modelled as a slab 6 layers} thick (left electrode). 
Next we were adding Al and O atoms and relaxed the geometry until we found a stable interface having 
two layers of oxygen atoms (2L). Finally we enclosed the interface with {\AF four ideal} Al(111) layers 
(right electrode) 
and connected it with the left electrode through periodic boundary {\AF conditions.}
Performing this procedure two different geometries of {\AF the} interfaces were identified: 
(1) {\AF an} asymmetric structure, corresponding the the ultra-thin {\AF AlO$_x$} 
layer investigated by Jennison, and (2) {\AF an} symmetric structure 
which did not contain the layer of chemisorbed oxygen next to the bottom Al electrode. 
More details on the differences between the asymmetric and symmetric structures can be found 
elsewhere~\cite{Dieskova07}; in our present work we will consider only structures derived 
from the asymmetric {\AF geometry}.

\begin{figure}
\begin{center}
       \includegraphics[width=8cm]{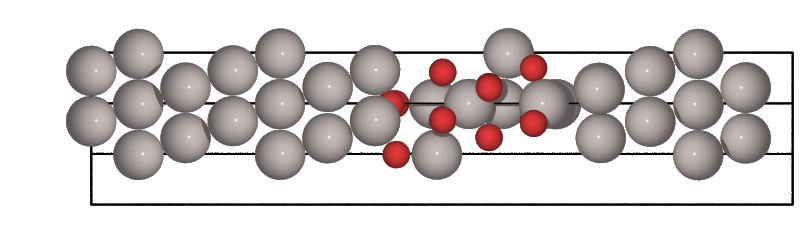}  \\
       \includegraphics[width=8cm]{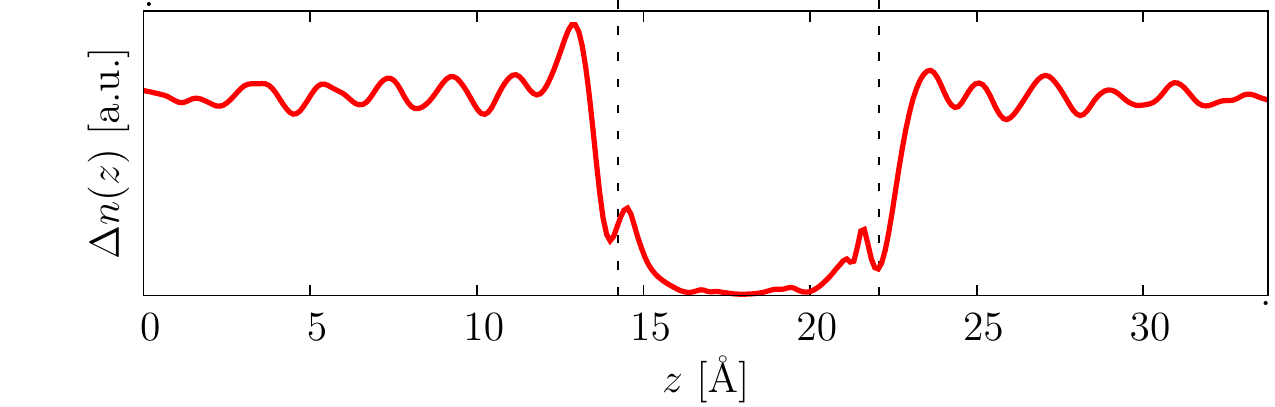}
\caption{(Color online) The 4L structure (above) and the corresponding averaged electronic density 
of the occupied transmitting states. The dashed lines indicate the positions of the metal-oxide boundary
obtained according to {\AF Eq.~(\ref{eq-4-1})}.
	} \label{fig-4-1}
\end{center}
\end{figure}
Motivated by the geometry of the asymmetric 2L interface model we have constructed thicker Al/AlO$_x$/Al 
by adding one (3L), two (4L) or three (5L) full oxygen layers sandwiched between monoatomic (Al1) 
or diatomic (Al2) {\AF layers} of aluminum. {\AF The resulting geometries} were optimized until the 
forces on the atoms were smaller than $0.002$ Ha/$a_B$, while the Al atoms beyond the first layer of bulk metal 
were kept fixed. An example of the resulting geometric structure of 4L is shown in Fig.~\ref{fig-4-1}.
We should mention that these models are not necessarily the only ones possible for the interface of the concerned 
width. Due to the above described tendency of AlO$_x$ systems towards {\AF disorder, we expect that many different 
variations could be found with larger surface cells. The structures identified here} 
need to be taken as few samples of the great variety of possible geometrical arrangements. However, 
the comparison of the projected density of states (PDOS) for symmetric and asymmetric 2L junctions 
{\AF(see Ref.~[\onlinecite{Dieskova07}])} {\AF suggests} that these differences lead to small changes 
in their conductances.

All of the ground state properties and optimizations were done using the \texttt{Quantum Espresso} 
{\AF distribution}\cite{Espresso}.
We have employed the PBE exchange-correlation functional, atomic cores were described using ultra-soft 
pseudopotentials resulting in well converged electronic structure 
close to the Fermi energy, using {\AF a cutoff energy of 12.5 (125) Ha for wavefunctions (charge density).}
Due to the large size of the supercell, {\AF a} $6\times6\times1$ Monkhorst-Pack k-point grid was 
sufficient {\AF to converge} the total energy and the electronic density.

In Fig.~\ref{fig-4-1}, in parallel with the geometric structure of the 4L interface, we show the profile of the 
plane-averaged electronic density of the scattering states $\Delta n(z)$ (e.g. localized states on oxygen atoms 
are not included). We see that the rapid drop and increase in the density appears at the boundary between 
the metal and the oxide.
We use $\Delta n(z)$ as the quantity for the determination of the interface width from our {\it ab initio} 
calculations, in close analogy with the determination of the position of surfaces at metal-vacuum 
{\AF interfaces}~\cite{Liebsch99}; for the left boundary we use 
\begin{equation}
	z_L = \int_{-\infty}^{z_I} z \D{\Delta n(z)}{z} dz / 
		\int_{-\infty}^{z_I} \D{\Delta n(z)}{z} dz, \label{eq-4-1}
\end{equation}
where $z_I$ is a position in the center of the insulator. Similar expression is used for the determination of the
right boundary $z_R$ which together with $z_L$ give the estimate of the interface width $d = z_R-z_L$ 
used within our potential barrier and $sp$ models in Sec.~\ref{sec-5}. The resulting {\AF interface}
widths are given in the Table~\ref{tab-1}. In {\AF the following} 
we will also refer to the width of the transition region 
beween the metal and the oxide, which can be estimated from the averaged density to be $\Delta n \approx 2.0$\AA.
This value will be used for the determination of the {\AF width} 
of the transition region in the potential barrier model (Fig.~\ref{fig-2-1}).
\begin{table}
\begin{center}
\begin{tabular}{c|c|c|c|c}
\hline \hline
System		& $d$ [\AA]	& $E_g$ [eV]	& $\Delta E_{F}$ [eV]	& $\nu(E_F)$	\\
\hline
2L		& $4.5$		& $7.0$		& $1.5$		& $0.82$	\\
3L		& $5.5$		& $6.5$		& $-0.25$	& $0.99$	\\
4L		& $7.8$		& $6.5$		& $-1.0$	& $0.91$	\\
5L		& $9.8$		& $6.5$		& $-1.0$	& $0.91$	\\
\hline \hline
\end{tabular}
\caption{Values of the interface widths, band-gaps, Fermi energy shifts and the shift factor $\nu(E_F)$ obtained 
from {\it ab initio} calculations.} \label{tab-1}
\end{center}
\end{table}

\begin{figure}
\begin{center}
       \includegraphics[width=8cm]{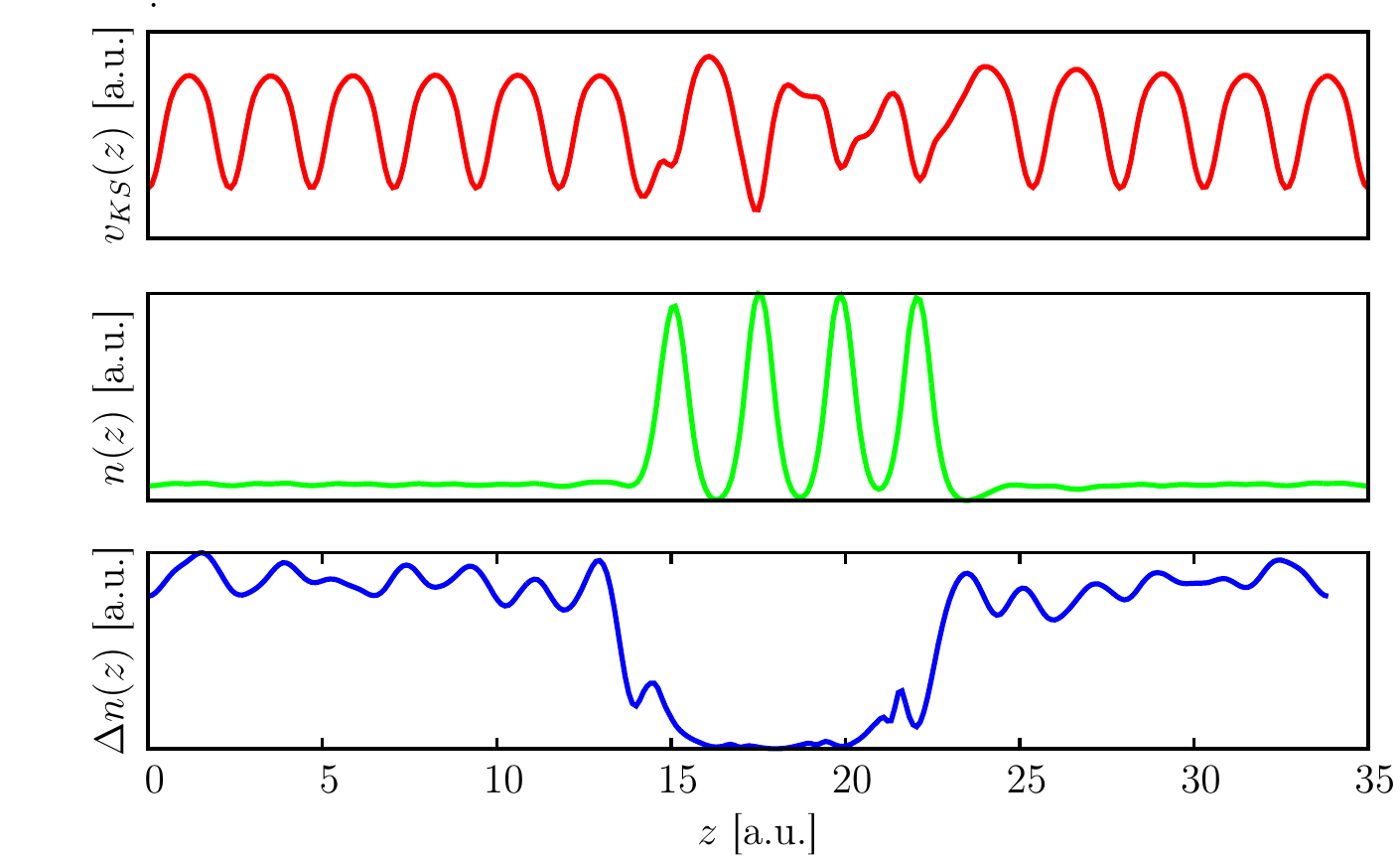}
\caption{(Color online) The comparison of the average Kohn-Sham potential $v_{KS}(x)$, the 
	total density $n(z)$ and the contribution to the density from the states close to the Fermi energy
	$\Delta n(z)$. Clearly, the latter can be unambiguously used for the definition of the 
	width of the interface using the Eq.~\ref{eq-4-1}.
	} \label{fig-4-2}
\end{center}
\end{figure}

{\AF We note} that for the calculation of the positions $z_{L/R}$ we do not {\AF necessarily} need to use the density 
$\Delta n(z)$ obtained from the scattering states in the transport calculations (see Sec~\ref{sec-5}), 
but is it equally good to use the partial density of the states close to the Fermi energy that can be obtained 
from any ground state code (e.g. \texttt{Quantum Espresso}). On the other hand, the total electronic density 
or the Kohn-Sham potential (which is frequently but incorrectly believed to be the origin of 
the potential barrier in the model from Sec.~\ref{sec-2}) are not suitable for this calculation, as it 
is clearly demonstrated in Fig.~\ref{fig-4-2}.

\begin{figure}
\begin{center}
       \includegraphics[width=8cm]{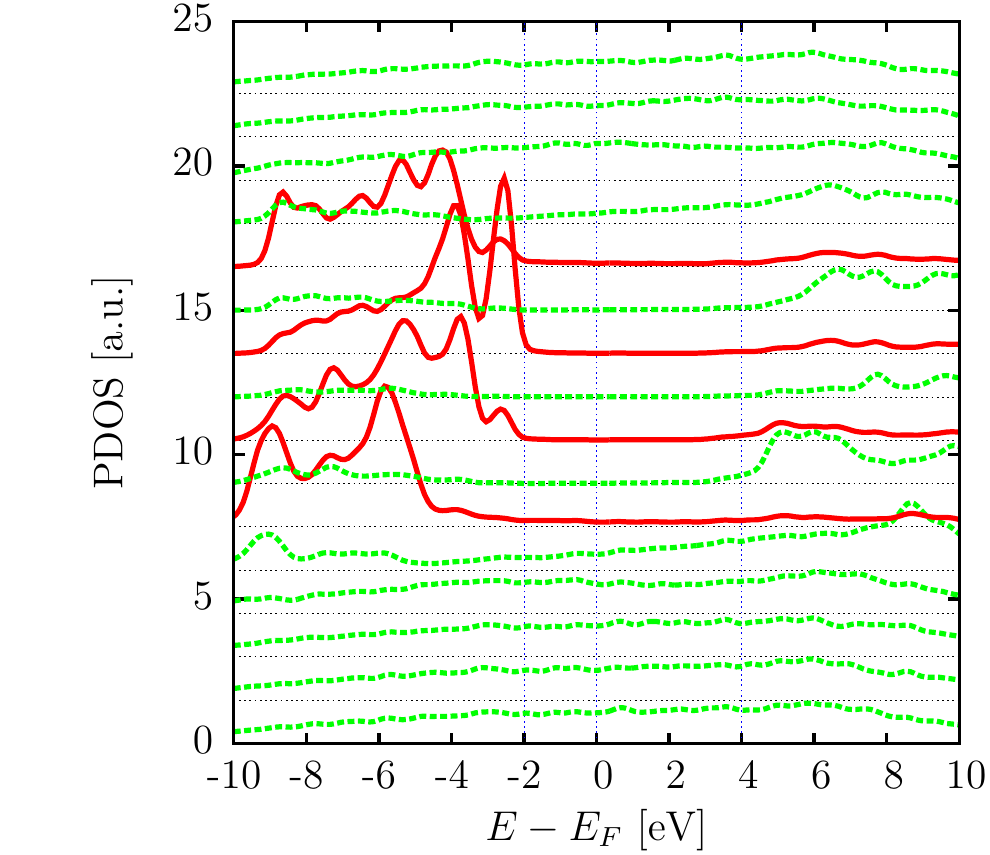}  \\
\caption{(Color online) The projected density of states for the AlOx 4L interface. The distance 
	of the valence band to the Fermi energy, being less than half the distance to the conduction band,
	is taken as the effective barrier height. 
	} \label{fig-4-3}
\end{center}
\end{figure}

The second important parameter of the potential barrier and the atomic $sp$ models is the {\AF insulator}
band gap $E_g$. {\AF It} can be extracted from the projected density of states (PDOS), where the Kohn-Sham 
eigenstates of the interface are projected {\AF on atomic} orbitals. Figure~\ref{fig-4-3} shows the PDOS 
for the 4L interface, where the PDOS of atoms in each layer are added together, green lines corresponding 
to the Al layers and the red lines to the oxygen layers. 
The {\AF oxide bandgap} is estimated {\AF as} the energy distance between the onset of the valence bands 
on the oxygen atoms below the Fermi energy, and the onset of the mixed Al and O {\AF bands} above the Fermi energy.
From the PDOS we can also obtain the energy distance between the midgap energy and the Fermi energy, 
$\Delta E_F$, needed for the $sp$ model. {\AF The calculated} bandgaps and $\Delta E_F$ for all studied 
interfaces are collected in the Table~\ref{tab-1}. Interestingly, in spite of the well known bandgap 
problem of the DFT~\cite{Sham1983,Godby1986,Perdew1983}, these bandgaps {\AF appear to be} in very good agreement 
with recent experimental results for the Al/Al$_2$O$_3$ interfaces\cite{Nguyen2008,Afanasev2002} 
which found $E_g=6.4$eV.

While the bandgap stays roughly the same for all of the studied interfaces $2L-5L$, the Fermi energy shifts 
with respect to the middle of the gap from positive (conventionally called the electron tunneling regime) 
to negative values (hole tunneling). However, the factor $\nu(E_F)$ {\AF stays} close to one 
in all the cases (see Table~\ref{tab-1}), as anticipated already in Sec~\ref{sec-2}.
The energy difference between the bottom of the conduction band and the Fermi energy determined 
experimentally~\cite{Nguyen2008} was found to be $E_c-E_F = 2.9 \pm 0.2$eV which is $1$eV smaller 
than the DFT value found here for $4L$ and $5L$, but on the other hand, in good agreement with $2L$ and $3L$, 
which perhaps indicates larger sensitivity of this quantity on the particular system.

\section{Electronic structure of an ideal insulator}
\label{sec-4b}

The potential barrier model as well as the $sp$ model also rely on the knowledge of the effective mass 
$m_{\AF\text{eff}}$
of the electrons in the insulator or barrier region. To calculate it we have considered {\AF a} first principles model
of the insulator extracted from the {\AF geometry of the 4L junction}.
Namely, it consists of a supercell of length $l=8.11 a_B$ in the $z$ direction 
and with identical dimensions in the two remaining in-plane directions {\AF [i.e. $\sqrt{3} \times \sqrt{3}$ Al(111)]}, 
The supercell contains two layers of oxygen and two layers of 2/3 filled Al planes. 
(the 3rd and 4th oxygen layers in Fig.~\ref{fig-4-1} from the left and their immediately following Al layers 
respectively). This way, the chemical composition actually corresponds to {\AF alumina, Al$_2$O$_3$}. 

The DFT ground state calculation has been done with the same specifications as for the full interface 
(Sec.~\ref{sec-4}) except for the k-point grid being here $6\times 6 \times 6$ due to smaller extent 
in the $z$ direction. The following band-structure calculation has been done using the \texttt{PWCOND}
program~\cite{Smogunov2004} that is capable of obtaining the so called complex band-structure, 
i.e. energy bands for imaginary as well as real Bloch $k$-vectors. We have checked that calculations 
of the band-structure for real $k$-vectors using the \texttt{Quantum Espresso} and the \texttt{PWCOND} 
gave identical results so that the parameters involved in the \texttt{PWCOND} program were correctly chosen. 

\begin{figure}
\begin{center}
       \includegraphics[width=8cm]{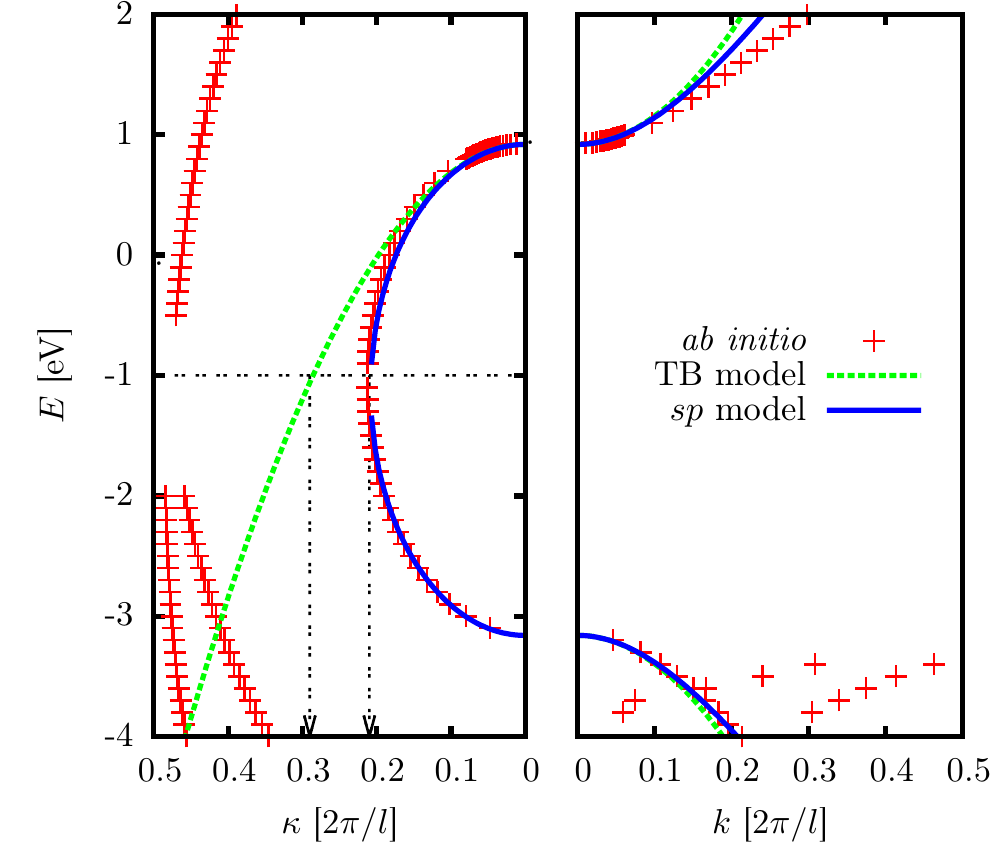}  \\
\caption{(Color online) The imaginary (left) and real (right) band structures from {\it ab inito} calculations compared with 
the band-structures of the $sp$ and potential barrier models. The $sp$ model gives an excellent fit 
for both real and imaginary band-structure for $m_{\text{eff}} = 0.35$ and $E_g=4.0$eV.
	} \label{fig-4b-1}
\end{center}
\end{figure}
The band-structure along the direction normal to the interface ($z$) is shown in Fig.~\ref{fig-4b-1}. First of all
we note that the bandgap obtained here, $E_g^\infty \sim 4$eV (in agreement with the previous DFT-PBE results for 
bulk $\gamma$-Al$_2$O$_3$\cite{Ahuja2004,Sankaran2010}), is significantly smaller that the bandgap extracted from 
the PDOS of the full junction ($\sim 6.5$eV). Interestingly, the experimental value of this phase of alumina 
is $E_g^{exp} = 7$eV, which can be obtained also computationally if the DFT-PBE result is followed by 
a GW calculation\cite{Sankaran2010}. 

The DFT band structure in Fig.~\ref{fig-4b-1} is fitted with two model dispersions. The TB model uses 
a free-electron like dispersion $E_c(k) = \epsilon_c + k^2/(2m_{\AF\text{eff}})$ which after fitting gives 
the effective mass $m_{\AF\text{eff}}=0.35$. The atomic $sp$ model {\AF [Eq.~(\ref{eq-3-1})]} 
in the approximation $ka/2<1$, 
which is used in the analytic {\AF expression} for the conductance, {\AF gives (for $k_x=k_y=0$) the dispersion:}
\begin{equation}
	E_{c/v}(k) = E_{F}^\infty \pm \frac{E_g}{2}\sqrt{ 1 + \frac{2 k_z^2}{m_{\AF\text{eff}} E_g} }.
\end{equation}
The parameters of the fit given {\AF in Fig.~\ref{fig-4b-1}} are $m_{\AF \text{eff}}=0.35$, $E_g=4.0$eV and 
$E_F^{\infty}=1$eV. Our value of the effective mass is to be compared with the electron's mass
obtained from DFT calculations for ideal $\alpha$-Al$_2$O$_3$ crystal, $m_{\AF\text{eff}} \approx 0.4$~\cite{Xu1991}, 
and fits to experimental $I-V$ characteristics, $m_{\AF\text{eff}} \approx 0.23$~\cite{Groner2002,Ganguly2011}.

We see that the $sp$ model works very well for real as well as imaginary band-structure
close to $k=0$. While both models give the same effective mass, the values of $\kappa$
for the free-electron like dispersion are larger by $\sim 50$\% (as indicated 
by arrows) which contributes to prediction of smaller conductances within the potential barrier 
model given the interface width is the same, as will be shown in the following section.

\section{The conductance}
\label{sec-5}

Transport properties of the junctions were obtained using the transfer matrix method~\cite{Joon1999} implemented 
in the \texttt{PWCOND} code~\cite{Smogunov2004}{\AF, using} plane-wave basis and ultra-soft pseudopotentials.
For the given self-consistent Kohn-Sham potential (obtained from the ground state calculations, see Sec.~\ref{sec-4}),
the {\AF conductance was converged with respect to} the $k_\parallel$ grid; going
from the $6\times6$ mesh (used for the presented results) to {\AF a} $10\times10$ mesh the change {\AF in} 
the conductance {\AF has been found to be} $<5$\%. 
Furthermore, for testing purposes, 
the conductances for {\AF the} $2L, 3L$ and $4L$ interfaces were also calculated 
using the \texttt{WanT} code~\cite{Ferretti05,Ferretti2007}, 
where {\AF a} completely different method based on maximally localized 
Wannier functions is {\AF implemented. Results are reported in Table~\ref{tab-2} and compare well
with the previous set, though slightly underestimating the absolute values.}

\begin{table}
\begin{center}
\begin{tabular}{c|c|c|c|c}
\hline \hline
    code    &    2L     &   3L         &   4L    &    5L   \\
\hline
 \texttt{PWCOND} &   $0.109$     &   $0.0166$   &   $0.00245$   &   $0.000279$  \\  
 \texttt{WanT}   &   $0.0668$    &   $0.00730$  &   $0.00224$   &   N/A         \\  
\hline \hline
\end{tabular}
\caption{Values of the conductances in multiplies of $e^2/h \times A_{z}$, where $A_{z}=74.23a_B^2$ 
is the area of the supercell perpendicular to the $z$ direction, calculated by the \texttt{PWCOND} 
and \texttt{WanT} codes. The differences are similar to the differences between the \texttt{PWCOND} results and the $sp$ model.}  \label{tab-2}
\end{center}
\end{table}

\begin{figure}
\begin{center}
       \includegraphics[width=8cm]{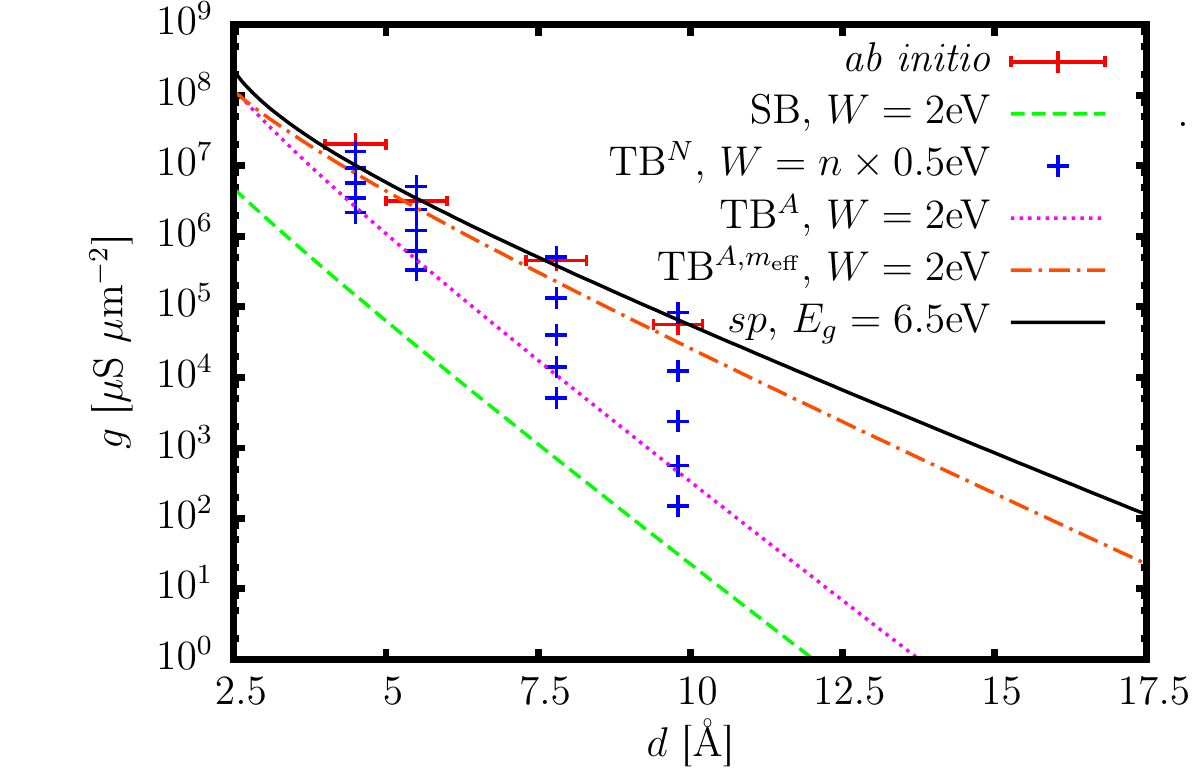}  \\
\caption{(Color online) The {\it ab initio} and model conductances. The square barrier (green dashed line),
the trapezoid barrier with transition regions $\Delta D \neq 0$ (pink dotted line) and the inclusion of the effective mass 
(orange dash-dotted line) form a sequence of improvements of the potential barrier to model towards the true 
atomic system. Finally, the analytic $sp$ model (black full line) with parameters taken from ground 
state {\it ab initio} simulations gives very good agreement with the full first principles calculation 
of the conductance.
	} \label{fig-5-1}
\end{center}
\end{figure}

{\AF In Figure}~\ref{fig-5-1} we show the dependence of the conductances per unit area on the interface
width $d$, determined in Sec.~\ref{sec-4}, in comparison with the two models considered in Sections~\ref{sec-2} 
and \ref{sec-3}. The horizontal error bars accompanying the {\it ab initio} conductances, $\Delta d \approx 2$\AA, 
indicate the width of the transition region between the metal and the insulator, which is taken from the 
averaged density profile, Fig.~\ref{fig-4-1}.  

First we consider the potential barrier model with effective mass equal to one, where the 
calculation of the transmission as well as its energy integration (Eq.~\ref{eq-2-2}) are done numerically 
exactly (TB$^{N}$). The potential barrier is of the form given in Fig.~\ref{fig-2-1}, where 
$d_w=d-\Delta d$. The conductances are shown as the blue crosses, where the height of the energy barrier 
$W=0.5n$eV, $n=1,2,3,4,5$ is increasing from top to bottom. The pink-dotted line is the conductance of the same 
potential barrier of width $W=2$eV, but evaluated using the approximate formula {\AF [Eq.~(\ref{eq-2-6})].}
As anticipated in Section~\ref{sec-2}, we see that in view of the overall differences, the approximate
but analytic formula is very satisfactory and the numerical calculation of the transmission of its energy integration
is not really needed. 

We see that in principle, we can achieve agreement between this model and the {\it ab initio} results 
if we choose $W\approx 0.5$eV, but this is in stark contrast with the estimates of the potential barrier 
height from the PDOS, typically taken as the distance between the Fermi energy and the nearest band 
in the oxide (e.g. the valence band in the oxide in 4L structure according to Fig.~\ref{fig-4-3}), 
here expected to be $W \sim 2$eV.

The green-dashed line is a conductance corresponding to a simple square potential barrier with $W=2$eV 
and effective mass equal to one, and we see that plain square barrier model goes in the wrong direction. 
The use transition regions of width $\Delta d$ does shift the potential barrier model in the right 
direction, particularly for very short interfaces, where the effective mass within the insulator does not 
seem to play an important role. Hence, use of the transition region between the metal and the insulator of width
$\Delta d$, given by the spatial extent of the drop if the electronic density between the metal and the oxide,
is essential for the TB model.

The red dash-dotted line gives the conductance {\AF according to Eq.~(\ref{eq-2-6}) } 
with the {\it ab initio} determined 
effective mass $m_{\AF\text{eff}}=0.35$ and $W=2$eV. The effective mass significantly improves the agreement
of the potential barrier model with the {\it ab initio} conductance, while keeping the barrier 
at the ``reasonable'' value, motivated by offset between the Fermi energy and the valence band maximum. 

Finally, the full black line corresponds to the atomic $sp$ model with the effective mass $m_{\AF \text{eff}}=0.35$,
band gap $E_g=6.5$eV and the barrier width $d_W = d-\Delta d$. The use of this reduced width 
$d_W$ is motivated by two {\AF observations:} (1) in Sec.~\ref{sec-2} we have seen {\AF that} 
the linearly increasing 
potential at distance $\Delta d$ contributes {\AF to the exponent of the conductance [Eqs.~(\ref{eq-2-6}-\ref{eq-2-7})]}
through a much smaller contribution $\Delta d_{E_{F}} = {\AF W/(W+E_F)} \Delta d \sim 0.15 \Delta d$. 
(2) in the TB model we have seen that {\AF the use of a} shorter barrier, 
effectively given by $d_W+2/3\Delta d$ {\AF [Eq.~(\ref{eq-2-7})]}, is important
to {\AF compare well} with the {\it ab initio} conductances. Hence we expect that also in the $sp$ model, 
the oxide width (i.e.  the equivalent of the potential {\AF barrier) needs} to be reduced almost to 
$d-\Delta d$, which is the value we use. 
As a result, the $sp$ model is essentially on top of the {\it ab initio} conductances. While the improvement 
with respect to the potential barrier model with transition region and the effective mass is not 
that large, it is important to {\AF stress that} the parameters of the $sp$ model ($E_g$, $m_{\AF \text{eff}}$, 
$d-\Delta d$) correspond to the characteristics of the true {\it ab initio} model.

\begin{figure}[!t]
\begin{center}
       \includegraphics[width=8cm]{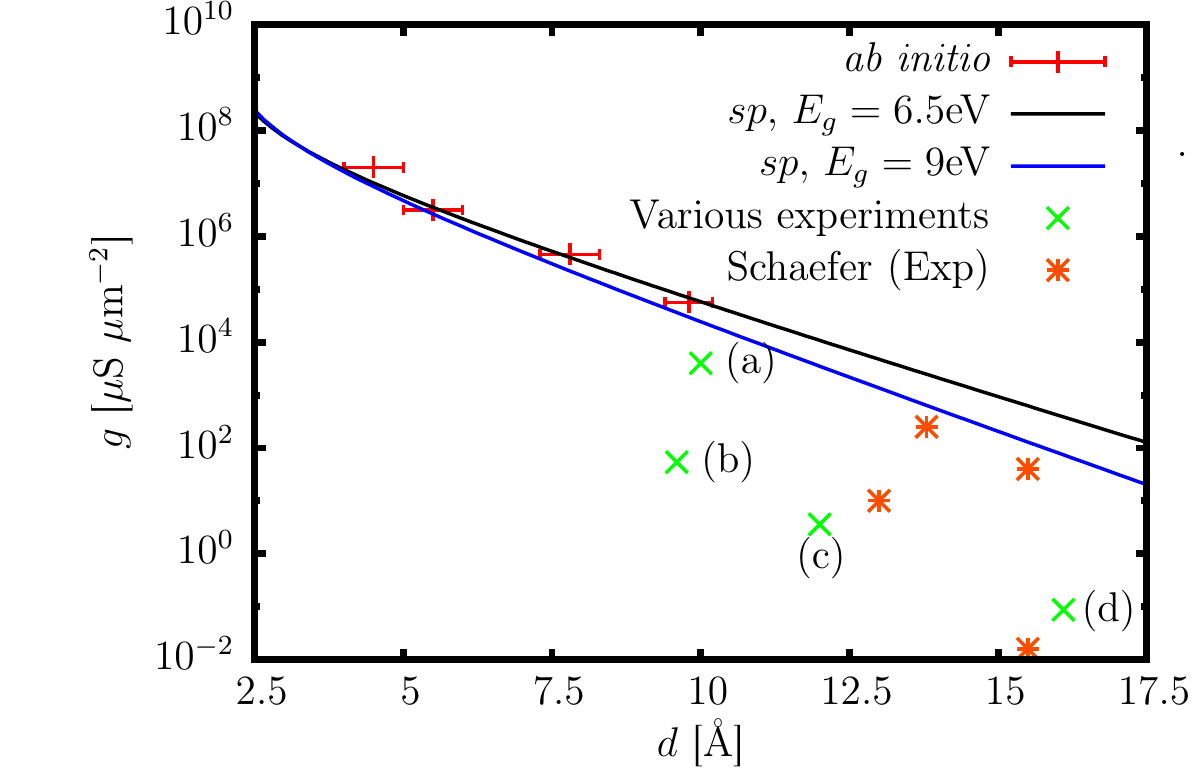}  \\
\caption{(Color online) The ab initio conductances compared to selected experimental results. (a) Jung~\cite{Jung09},
(b) Gloos~\cite{Gloos03}, (c) Holmqvist~\cite{Holmqvist2008}, (d) Brinkman~\cite{Brinkman1970}, and recent experiments 
by Schaefer~\cite{Schaefer2011}. The model gives fairly rigid prediction of the conductances, even 
using the bandgap of the $\alpha$-Al$_2$O$_3$, $E_g \sim 9$eV. The likely source of these discrepancies
is the experimental determination of the interface width.
	} \label{fig-5-2}
\end{center}
\end{figure}

It is interesting to attempt a quantitative comparison between experimentally determined barrier widths and heights,
and our {\it ab initio} {\AF and} $sp$ model results (Fig.~\ref{fig-5-2}). 
As mentioned already in the introduction, there are 
experimental junctions that are now accessible {\AF to} first principles simulations. Based on the rather unsatisfactory 
state of affairs in Fig.~\ref{fig-5-2} we suspect that not all of the 
{\AF published widths may have been determined correctly.}
On the other hand, a positive example is the data point taken from 
the work of Jung~\cite{Jung09}, where the interface width has been determined directly, and not through fits to 
the Simmons model and as a result the conductance is relatively closer our {\it ab initio} conductances. 
Similar underestimation of the junction widths obtained from Simmon's (potential barrier model) has been obtained 
in the experimental work of Buchanan {\it et al.}\cite{Buchanan2002}, even though here it has been interpreted 
as due to interface roughness.

\section{Conclusions}
\label{sec-7} 

In the conclusions, we have analyzed the performance of simple analytical models {\AF in describing the} conductance
of ultra-thin Al/AlO$_x$/Al {\AF junctions. We have compared} atomistic first-principles {\AF calculations
using the DFT-PBE framework} combined with the Landauer {\AF formula,} with the conductances obtained 
from {\AF a} potential barrier and {\AF a} tight-binding $sp$ analytical models. 
We have shown that {\AF the expression} for the conductance of the atomic $sp$ 
model has the same form as {\AF that from} the potential barrier model if the barrier height $W$ is exchanged for 
$\nu(E_F) E_g/4$ with $\nu(E_F) \sim 1$, which explains the small values of $W$ obtained frequently in the past 
by fitting the potential barrier model to the experimental $I-V$ curves. 
The {\AF accuracy} of the analytical models {\AF has been} tested by using parameters derived 
from ground-state DFT calculations. 
We have found that the oxide is characterized by {\AF effective mass $m_{\AF\text{eff}}=0.35$ and bandgap $E_g=6.5$eV.} 
{\AF When these parameters are used in combination with the $sp$ model, excellent agreement 
with the numerically calculated conductances is found}. 
The interface width used in the models has been shown to correspond to the width of 
the well-developed oxide which is shorter by about $\Delta d \approx 2$\AA \ compared to the geometric width 
of the interface $d$. 

\acknowledgements
This research has been supported by the Slovak Research and Development Agency under the contract 
{\AF No.~APVV-0108-11, and}
the Project HPC-EUROPA++ (RII3-CT-2003-506079).
PB would like to {\AF thank Kurt Gloos} and Hyunsik Im for the correspondence regarding their experimental data.
{\AF AF ackwnoledges support from Italian MIUR through
Grant No. FIRB-RBFR08FOAL\_001. }

\bibliography{References}

\begin{thebibliography}{10}

\bibitem{Fisher1961}
J.~C. Fisher and I. Giaever, J. Appl. Phys. {\bf 32},  172  (1961).

\bibitem{Holm1951}
R. Holm, J. Appl. Phys. {\bf 22},  569  (1951).

\bibitem{Stratton62}
R. Stratton, J. Phys. Chem. Solids {\bf 23},  1177  (1962).

\bibitem{Simmons1963}
J.~G. Simmons, J. Appl. Phys. {\bf 34},  1793  (1963).

\bibitem{Brinkman1970}
W.~F. Brinkman, R.~C. Dynes, and J.~M. Rowell, J. Appl. Phys. {\bf 41},  1915
  (1970).

\bibitem{Groner2002}
M. Groner, J. Elam, F. Fabreguette, and S. George, Thin Solid Films {\bf 413},
  186  (2002).

\bibitem{Buchanan2002}
J.~D.~R. Buchanan, T.~P.~A. Hase, B.~K. Tanner, N.~D. Hughes, and R.~J. Hicken,
  Appl. Phys. Lett. {\bf 81},  751  (2002).

\bibitem{Gloos03}
K. Gloos, P.~J. Koppinen, and J.~P. Pekola, J. of Phys. Cond. Matt. {\bf 15},
  1733  (2003).

\bibitem{Schaefer2011}
D.~M. Schaefer, P.~F.~P. Fichtner, M. Carara, L.~F. Schelp, and L.~S. Dorneles,
  Journal of Physics D: Applied Physics {\bf 44},  135403  (2011).

\bibitem{Miller2007}
C.~W. Miller, Z.-P. Li, J. Akerman, and I.~K. Schuller, Appl. Phys. Lett. {\bf
  90},  043513  (2007).

\bibitem{Lacquaniti2012}
V. Lacquaniti, M. Belogolovskii, C. Cassiago, N.~D. Leo, M. Fretto, and A.
  Sosso, New Journal of Physics {\bf 14},  023025  (2012).

\bibitem{Jonson80}
M. Jonson, Solid State Commun. {\bf 33},  743  (1980).

\bibitem{Bokes11}
P. Bokes, Phys. Rev. A {\bf 83},  032104  (2011).

\bibitem{Feibelman07}
P.~J. Feibelman, Phys. Rev. B {\bf 76},  235405  (2007).

\bibitem{Jung09}
H. Jung, Y. Kim, K. Jung, H. Im, Y.~A. Pashkin, O. Astafiev, Y. Nakamura, H.
  Lee, Y. Miyamoto, and J.~S. Tsai, Phys. Rev. B {\bf 80},  125413  (2009).

\bibitem{Fadlallah09}
M.~M. Fadlallah, C. Schuster, U. Schwingenschl\"ogl, I. Rungger, and U. Eckern,
  Phys. Rev. B {\bf 80},  235332  (2009).

\bibitem{Stoeffler02}
D. Stoeffler, EPL (Europhysics Letters) {\bf 59},  742  (2002).

\bibitem{Belashchenko05}
K.~D. Belashchenko, E.~Y. Tsymbal, I.~I. Oleynik, and M. van Schilfgaarde,
  Phys. Rev. B {\bf 71},  224422  (2005).

\bibitem{Zhuravlev11}
M.~Y. Zhuravlev, R.~F. Sabirianov, S.~S. Jaswal, and E.~Y. Tsymbal, Phys. Rev.
  Lett. {\bf 94},  246802  (2005).

\bibitem{Rippard02}
W.~H. Rippard, A.~C. Perrella, F.~J. Albert, and R.~A. Buhrman, Phys. Rev.
  Lett. {\bf 88},  046805  (2002).

\bibitem{Tan05}
E. Tan, P.~G. Mather, A.~C. Perrella, J.~C. Read, and R.~A. Buhrman, Phys. Rev.
  B {\bf 71},  161401  (2005).

\bibitem{Koppinen2007}
P.~J. Koppinen, L.~M. Vaisto, and I.~J. Maasilta, Appl. Phys. Lett. {\bf 90},
  053503  (2007).

\bibitem{Smogunov2004}
A. Smogunov, A. Dal~Corso, and E. Tosatti, Phys. Rev. B {\bf 70},  045417
  (2004).

\bibitem{Ferretti05}
A. Ferretti, A. Calzolari, R.~D. Felice, F. Manghi, M.~J. Caldas, M.~B.
  Nardelli, and E. Molinari, Phys. Rev. Lett. {\bf 94},  116802  (2005).

\bibitem{Ferretti2007}
A. Ferretti, A. Calzolari, B. Bonferroni, and R.~D. Felice, Journal of Physics:
  Condensed Matter {\bf 19},  036215  (2007).

\bibitem{Gundlach1973}
K.~H. Gundlach, J. Appl. Phys. {\bf 44},  5005  (1973).

\bibitem{Tomfohr2002}
J.~K. Tomfohr and O.~F. Sankey, Phys. Rev. B {\bf 65},  245105  (2002).

\bibitem{Prodan07}
E. Prodan and R. Car, Phys. Rev. B {\bf 76},  115102  (2007).

\bibitem{Ferretti2012}
A. Ferretti, G. Mallia, L. Martin-Samos, G. Bussi, A. Ruini, B. Montanari, and
  N.~M. Harrison, Phys. Rev. B {\bf 85},  235105  (2012).

\bibitem{Mizuguchi2005}
M. Mizuguchi, Y. Suzuki, T. Nagahama, and S. Yuasa, Appl. Phys. Lett. {\bf 87},
   171909  (2005).

\bibitem{Chen08}
M.~S. Chen and D.~W. Goodman, J. Phys.: Condens. Matter {\bf 20},  264013
  (2008).

\bibitem{Kravchuk04}
T. Kravchuk, R. Akhvlediani, V.~V. Gridin, and A. Hoffman, Surf. Sci. {\bf
  562},  83  (2004).

\bibitem{Nesbitt2007}
J.~R. Nesbitt and A.~F. Hebard, Phys. Rev. B {\bf 75},  195441  (2007).

\bibitem{Hasnaoui05}
A. Hasnaoui, O. Politano, J.~M. Salazar, G. Aral, R.~K. Kalia, A. Nakano, and
  P. Vashishta, Surf. Sci. {\bf 579},  47  (2005).

\bibitem{Jennison1999}
D.~R. Jennison, C. Verdozzi, P.~A. Schultz, and M.~P. Sears, Phys. Rev. B {\bf
  59},  R15605  (1999).

\bibitem{Jennison2000}
D. Jennison and A. Bogicevic, Surface Science {\bf 464},  108  (2000).

\bibitem{Dieskova07}
M. Dieskova, M. Konopka, and P. Bokes, Surf. Science {\bf 601},  4134  (2007).

\bibitem{Espresso}
P. Giannozzi, S. Baroni, N. Bonini, M. Calandra, R. Car, C. Cavazzoni, D.
  Ceresoli, G.~L. Chiarotti, M. Cococcioni, I. Dabo, A.~D. Corso, S. Fabris, G.
  Fratesi, S. de~Gironcoli, R. Gebauer, U. Gerstmann, C. Gougoussis, A. Kokalj,
  M. Lazzeri, L. Martin-Samos, N. Marzari, F. Mauri, R. Mazzarello, S. Paolini,
  A. Pasquarello, L. Paulatto, C. Sbraccia, S. Scandolo, G. Sclauzero, A.~P.
  Seitsonen, A. Smogunov, P. Umari, and R.~M. Wentzcovitch, J. Phys. Condens.
  Matter {\bf 21},  395502  (2009).

\bibitem{Liebsch99}
A. Liebsch, {\em Electronic Excitations at Metal Surfaces} (Plenum Press, New
  York, 1997).

\bibitem{Sham1983}
L.~J. Sham and M. Schluter, Phys. Rev. Lett. {\bf 51},  1888  (1983).

\bibitem{Godby1986}
R.~W. Godby, M. Schluter, and L.~J. Sham, Phys. Rev. Lett. {\bf 56},  2415
  (1986).

\bibitem{Perdew1983}
J.~P. Perdew and M. Levy, Phys. Rev. Lett. {\bf 51},  1884  (1983).

\bibitem{Nguyen2008}
N.~V. Nguyen, O.~A. Kirillov, W. Jiang, W. Wang, J.~S. Suehle, P.~D. Ye, Y.
  Xuan, N. Goel, K.-W. Choi, W. Tsai, and S. Sayan, Appl. Phys. Lett. {\bf 93},
   082105  (2008).

\bibitem{Afanasev2002}
V.~V. Afanas'ev, M. Houssa, A. Stesmans, and M.~M. Heyns, J. Appl. Phys. {\bf
  91},  3079  (2002).

\bibitem{Ahuja2004}
R. Ahuja, J.~M. Osorio-Guillen, J.~S. de~Almeida, B. Holm, W.~Y. Ching, and B.
  Johansson, Journal of Physics: Condensed Matter {\bf 16},  2891  (2004).

\bibitem{Sankaran2010}
K. Sankaran, G. Pourtois, R. Degraeve, M.~B. Zahid, G.-M. Rignanese, and J.
  Van~Houdt, Appl. Phys. Lett. {\bf 97},  212906  (2010).

\bibitem{Xu1991}
Y.-N. Xu and W.~Y. Ching, Phys. Rev. B {\bf 43},  4461  (1991).

\bibitem{Ganguly2011}
S. Ganguly, J. Verma, G. Li, T. Zimmermann, H. Xing, and D. Jena,  in {\em
  Device Research Conference (DRC), 2011 69th Annual} (IEEE, Santa Barbara,
  2011), pp.\ 121--122.

\bibitem{Joon1999}
H. J. Choi and J. Ihm, Phys. Rev. B {\bf 59},  2267  (1999).

\bibitem{Holmqvist2008}
T. Holmqvist, M. Meschke, and J.~P. Pekola, J. Vac. Sci. Technol. B {\bf 26},
  28  (2008).

\bibitem{Franz56}
W. Franz,  in {\em Handbuch der Physik}, edited by S. Fluegge (Springer,
  Berlin, 1956), p.\ 155.

\end{thebibliography}
\bibliographystyle{prsty}

\end{document}